\documentclass[pra,showkeys,preprint]{revtex4}

\usepackage{graphicx,booktabs,array}
\usepackage{amsmath}
\usepackage{amssymb}

\usepackage{graphicx,mathdots}
\usepackage{booktabs}
\usepackage{mathtools,tabularx}
\usepackage{makecell}
\setcellgapes{3pt}
\usepackage{graphicx,color}

\usepackage{graphicx,amssymb}
\usepackage{graphicx}

\begin{document}
\setcounter{page}{1}


\title{Exact solution of the Klein–Gordon equation in a Kiselev black hole background}

\author{M. D. de Oliveira$^{1}$\footnote{Corresponding author. Email: dalpra.matheus@gmail.com} and Alexandre G. M. Schmidt$^{1}$}
\affiliation{Instituto de Ci\^encias Exatas, Universidade Federal Fluminense,\\ 
27213-145 Volta Redonda --- RJ, Brazil}


\begin{abstract}

In this work, we investigate exactly the dynamics of a relativistic spinless particle influenced by a static and uncharged black hole surrounded by a quintessence-like anisotropic fluid or Kiselev black hole. Considering two quintessence-like models as examples, we calculate the radial wave function and determine, in both cases, the quasispectrum of energy, as well as the Hawking radiation and temperature. We find that the stronger the influence of quintessence-like anisotropic fluid, the smaller the radiation observed outside the event horizon. Furthermore, the temperatures obtained in both cases are directly influenced by the quintessence-like anisotropic fluid, and we recover the values obtained in other contexts, such as those derived using the surface gravity. Finally, in the absence of quintessence, we recover in both cases the Hawking temperature $T_{H} = 1/(8\pi k_{B} M)$ corresponding to the Schwarzschild black hole.

\end{abstract}

\maketitle

\section{Introduction}

The investigation of compact gravitational objects, particularly black holes, constitutes one of the pillars of modern theoretical physics, establishing deep connections between general relativity, quantum field theory, and cosmology. Classical solutions of Einstein’s equations, such as the Schwarszchild and Kerr spacetimes \cite{kerr,gravitation}, have been widely used as fundamental scenarios to explore extreme gravitational effects. However, observational evidence indicates that the universe is dominated by a dark energy component responsible for the accelerated expansion of the cosmos \cite{riess,perlmutter,planck}, whose nature remains an open question, motivating the analysis of its effects on black hole geometries.

Among the models proposed to describe dark energy, quintessence stands out, being characterized by a dynamical scalar field with negative pressure \cite{ratra,caldwell,tsujikawa}. Unlike the cosmological constant, quintessence allows for a more general and dynamic description of cosmic acceleration. In this context, the model introduced by Kiselev \cite{kiselev} provides an exact solution of Einstein’s equations describing a black hole surrounded by quintessence, in which the metric is modified by a term depending on a parameter associated with the dark energy density. However, as definitively established by Matt Visser \cite{visser}, the Kiselev spacetime is intrinsically anisotropic and cannot be classified as a perfect fluid, meaning it does not describe standard cosmological quintessence. Thus, the geometry of this type of spacetime model serves as an anisotropic toy model rather than a realistic representation of cosmic dark energy, and it may be described as a spacetime surrounded by a quintessence-like anisotropic fluid, or simply as a Kiselev black hole. Several studies have shown that this quintessence-like anisotropic fluid can significantly modify important physical quantities, such as the Hawking temperature, entropy, and thermodynamic stability of these systems \cite{zhou,liu,ma}. Moreover, recent analysis indicate that the presence of dark energy can affect astrophysical observables, such as black hole shadows, photon orbits, and quasinormal modes \cite{tsupko,amarilla,wei}. In particular, the investigation of quasinormal modes has proven to be an important tool for characterizing the dynamical response of black holes under perturbations \cite{konoplya0,berti}, revealing that quintessence directly influences both the oscillation frequencies and the decay rates of these modes.

In parallel, the study of quantum fields in curved spacetimes has revealed that black holes in different contexts \cite{suzuki,hatsuda,bezerra1,vieira, vieira-2015, vieira-class-quantum-grav,hortacsu, Senjaya1,Senjaya2,Senjaya3,Senjaya4,Senjaya5,Senjaya6,Senjaya7,Senjaya8,Senjaya9,kokkotas} emit thermal radiation due to quantum effects near the event horizon, a phenomenon known as Hawking radiation \cite{hawking,bekenstein}. These results highlight the importance of the cosmological environment in the description of quantum processes in strong gravitational regimes.

Motivated by these advances, the analysis of the dynamics of quantum particles has been widely used to investigate how the presence of dark energy modifies the emission spectrum, decay rates, and fundamental properties of black holes, including horizon structure, thermodynamic properties, and geodesic trajectories \cite{saleh,ghoshhawking,chen,jing,ghosh,fernando}. In this work, we investigate the dynamics of relativistic spinless quantum particles in black hole spacetimes surrounded by a quintessence-like anisotropic fluid or Kiselev black hole, with emphasis on determining the quasispectrum energy and analyzing Hawking radiation through the radial wave function. Considering two different quintessence-like models, as will be discussed in the next section, we analyze how distinct manifestations of dark energy influence both the quantum behavior of particles and the thermodynamic properties of the black hole. Such analysis may provide relevant insights into the interaction between quantum matter and spacetime geometry, especially in the vicinity of the event horizon.

The outline of this paper is as follows: In Section II, we present the two types of quintessence models that will be investigated, and we briefly review the formalism necessary to determine the Klein–Gordon equation in a Kiselev black hole background. In the same section, we decouple the angular and radial equations and solve the angular equation. In Section III, we solve the radial equation for both cases, obtain the quasi-energy spectrum, and analyze the Hawking radiation and Hawking temperature at the event horizon. In this section, we also examine, through graphical analysis, how dark energy influences the radiation observed in the region outside the event horizon. Finally, in Section IV, we present our conclusions.

\section{Spinless particle in the Kiselev black hole spacetime}

Let the metric of the static uncharged Kiselev black hole be given by \cite{kiselev}
\begin{equation}\label{metricageral}
	ds^2 = \left(1-\frac{2M}{r} - \frac{\alpha}{r^{3\omega_{0} + 1}}\right)dt^2 - \left(1-\frac{2M}{r} - \frac{\alpha}{r^{3\omega_{0} + 1}}\right)^{-1}dr^2 - r^2 d\theta^2 -r^2\sin^{2}\theta d\phi^2,
\end{equation}
where $M$ is the mass of the black hole, $-1 < \omega_{0} \leq -1/3$, and $\alpha$ is a constant associated with dark energy (quintessence-like). In this work, we will investigate two different values of $\omega_{0}$. The first case is $\omega_{0} = -1/3$, thus we have
\begin{equation}
	1-\frac{2M}{r} - \frac{\alpha}{r^{3\omega_{0} + 1}} = 1-\alpha - \frac{2M}{r},
\end{equation}
for this case, the event horizon will be given by
\begin{equation}
	1-\alpha - \frac{2M}{r_{h}} = 0 \hspace{0.5cm} \rightarrow \hspace{0.5cm} r_{h} = \frac{2M}{1-\alpha},
\end{equation}
in the limit $\alpha \rightarrow 0$, we recover the value $r_{h} = 2M$ of the Schwarszchild black hole. Furthermore, for $r_h$ to be non-negative, it is necessary that $\alpha < 1$. The second case is $\omega_{0} = -2/3$, thus we have
\begin{equation}
	1-\frac{2M}{r} - \frac{\alpha}{r^{3\omega_{0} + 1}} = -\frac{(2M -r + \alpha r^2)}{r} ,
\end{equation}
in this case, the event horizons will be given by
\begin{equation}
	\alpha r^2 - r + 2M = 0 \hspace{0.5cm}\rightarrow \hspace{0.5cm} r_{\pm} = \frac{1}{2\alpha} \pm \frac{1}{2\alpha}\sqrt{1-8\alpha M},
\end{equation}
where the event horizon is $r_{h} = r_{-}$, since in the limit $\alpha \rightarrow 0$, we recover the value $r_{h} = 2M$ and $r_{+}$ is commonly called as the quintessence-like horizon. Thus, we observe that in two cases the condition for the physical event horizon is the one in which, in the limit $\alpha \rightarrow 0$, we recover the event horizon of the Schwarszchild black hole, $r_{h} = 2M$. Furthermore, for the horizons to be real, the condition $\alpha \leq 1/8M$ must be satisfied. Otherwise, the black hole will not possess any horizons, resulting in a naked singularity at $r \rightarrow 0.$

In this work, we will investigate the behavior of a relativistic spinless test particle. For this purpose, we begin by considering the Klein–Gordon equation with $\hbar = c = 1$, for a generalized metric $ds^2 = g_{\mu\nu}dx^{\mu}dx^{\nu}$ which is given by
\begin{equation}\label{kgmetricacurva}
\frac{1}{\sqrt{-g}}\partial_{\mu}(\sqrt{-g}g^{\mu\nu}\partial_{\nu}\Psi) = - m_{0}^{2}\Psi
\end{equation}
where $m_{0}$ is the mass of the scalar particle, with  $g = {\rm det}(g_{\mu\nu})$ where $g_{\mu \nu}$ and $g^{\mu \nu}$ are the metric tensor and its inverse, respectively, and we use the Einstein convention sum with repeated Greek indices run from 0 to 3. From (\ref{metricageral}) we identify the metric tensor $g_{\mu \nu}$,
\begin{equation}\label{tensormetrico}
g_{\mu\nu}= {\rm diag}\left[1-\frac{2M}{r} - \frac{\alpha}{r^{3\omega_{0} + 1}},-\left(1-\frac{2M}{r} - \frac{\alpha}{r^{3\omega_{0} + 1}}\right)^{-1},-r^2,-r^2\sin^2\theta\right],
\end{equation}
thus the inverse metric tensor is,
\begin{equation}
g^{\mu\nu} = {\rm diag}\left[\left(1-\frac{2M}{r} - \frac{\alpha}{r^{3\omega_{0} + 1}}\right)^{-1},-\left(1-\frac{2M}{r} - \frac{\alpha}{r^{3\omega_{0} + 1}}\right),-\frac{1}{r^2},-\frac{1}{r^2\sin^2\theta}\right].
\end{equation}

Using $g = -r^4\sin^2(\theta)$ and (\ref{tensormetrico}), we write the Klein-Gordon equation in a Kiselev black hole background as, 
\begin{eqnarray}\label{kgequationfinal}
\left\{h^{2}(r)\left[\frac{\partial^2}{\partial r^2} + \left(\frac{2}{r} + \frac{h^{'}(r)}{h(r)}\right)\frac{\partial}{\partial r}\right] + \frac{h(r)}{r^2} \hat{L}^{2} - m_{0}^{2} h(r)- \frac{\partial^2}{\partial t^2}\right\}\Psi  = 0,
\end{eqnarray}
where $h(r) = 1 - 2M/r - \alpha r^{-3\omega_{0} - 1}$ with $h^{'}(r)$ is its the first derivative, and the square angular momentum operator $\hat{L}^{2}$ is given by
\begin{equation}\label{angularoperator}
\hat{L}^{2} = \frac{\partial^2}{\partial \theta^2} + \cot(\theta)\frac{\partial}{\partial \theta} + \frac{1}{\sin^2(\theta)} \frac{\partial^2}{\partial \phi^2},
\end{equation}
and its eigenfunctions are the well-known spherical harmonics $Y^{m}_{l}(\theta,\phi)$ \cite{arfken}, with $\hat{L}^{2} Y^{m}_{l}(\theta,\phi) = -l(l+1) Y^{m}_{l}(\theta,\phi)$, where $l$ is the orbital angular momentum and $m$ is the quantum number associated with the magnetic moment. Thus, due to the spherical symmetry of the problem, we can write $\Psi(r,\theta,\phi,t) = R(r) Y^{m}_{l}(\theta,\phi) e^{-i \omega t}$, with $\omega$ being the energy, and by substituting this into (\ref{kgequationfinal}) we obtain the radial equation.
\begin{eqnarray}\label{radialequationgeral}
	\left\{h^{2}(r)\left[\frac{d^2}{d r^2} + \left(\frac{2}{r} + \frac{h^{'}(r)}{h(r)}\right)\frac{d}{d r}\right] - l(l+1) \frac{h(r)}{r^2}  - m_{0}^{2} h(r) + \omega^2\right\}R(r)  = 0.
\end{eqnarray}

Thus, by defining the value of $\omega_{0}$, we obtain the form of $h(r)$ and can solve the radial equation. In the next section, we will analyze the two cases considered above for $\omega_{0}$ and investigate the physical consequences of the different manifestations of dark energy on the dynamics of a quantum particle.

\section{Radial wave equation exact solution}

In this section, we will calculate the value of $R(r)$ for two different values of $\omega$ and investigate the quasispectrum of energy, as well as the decay rate, Hawking radiation, and temperature through the radial wave function in the region of the event horizon of the black hole.

\subsection{First case: $\omega_{0} = -1/3$}

In this first case, we consider $\omega_{0} = -1/3$. Thus, using (\ref{radialequationgeral}), we have
\begin{eqnarray}\label{radialequation1}
	\left[\frac{d^2}{d r^2} + \left(\frac{2}{r} + \frac{2M}{r[(1-\alpha)r - 2M]}\right)\frac{d}{d r} - \frac{l(l+1)}{r[(1-\alpha)r - 2M]} - m_{0}^{2} \frac{r}{(1-\alpha)r - 2M} \nonumber \right. \\\\ \left.  + \omega^2 \frac{r^2}{[(1-\alpha)r - 2M]^2}\right]R(r)  = 0.\nonumber
\end{eqnarray}

To solve the differential equation above, we first make the substitution $x = 1 - r/r_{h}$, where $r_{h} = 2M/(1-\alpha)$, thus we obtain
\begin{eqnarray}\label{radialequation1x}
	\left\{\frac{d^2}{d x^2} + \left(\frac{1}{x} + \frac{1}{x-1}\right)\frac{d}{d x} + \left[\frac{l(l+1)}{1-\alpha} + \frac{4M^2 m_{0}^{2}}{(1-\alpha)^3} - \frac{8M^2 \omega^2}{(1-\alpha)^4}\right]\frac{1}{x} + \frac{4M^2 \omega^2}{(1-\alpha)^4}\frac{1}{x^2} \nonumber \right. \\\\ \left. -\frac{l(l+1)}{(1-\alpha)}\frac{1}{x-1}- \frac{4m_{0}^{2} M^2}{(1-\alpha)^3} + \frac{4M^2 \omega^2}{(1-\alpha)^4}\right\}R(x)  = 0,\nonumber
\end{eqnarray}
the solution of this differential equation can be obtained in terms of the confluent Heun function via the ansatz $R(x) = x^{s_{1}}e^{s_{2}x} f(x)$. In this way, we eliminate the constant and $x^{-2}$ terms from the equation, and thus we obtain
\begin{eqnarray}\label{radialequation1f}
	\left\{\frac{d^2}{d x^2} + \left(\frac{2s_{1}+1}{x} + \frac{1}{x-1} + 2s_{2}\right)\frac{d}{d x} + \frac{A}{x} + \frac{B}{x-1}\right\}f(x)  = 0,\nonumber
\end{eqnarray}
where $A = l(l+1)/(1-\alpha) + 4M^2 m_{0}^{2}/(1-\alpha)^3 - 8M^2 \omega^2/(1-\alpha)^4 - s_{1} + s_{2} + 2s_{1}s_{2}$ and $B = -l(l+1)/(1-\alpha) + s_{1} + s_{2}$, $s_{1} = \pm 2iM\omega/(1-\alpha)^2$ and $s_{2} = \pm 2M \sqrt{m_{0}^{2}(1-\alpha)-\omega^2}/(1-\alpha)^2$. The equation (\ref{radialequation1f}) is analogous to the confluent Heun equation, which has two finite regular singularities at $z = (0,1)$ and one irregular singularity at infinity $z \rightarrow \infty$, and in canonical form is given by \cite{ronveaux}
\begin{equation}
	y''(z) + \left(\frac{\gamma}{z} + \frac{\delta}{z-1} + \epsilon\right)y'(z) + \frac{(\eta z - q)}{z(z-1)}y(z) = 0,
\end{equation}
whose solution is
 \begin{equation}
 	y(z) = C_{1} H_{c}(q,\eta,\gamma,\delta,\epsilon;z) + C_{2} z^{1-\gamma}H_{c}(q',\eta',\gamma',\delta,\epsilon;z),
 \end{equation}
where $H_{c}$ is the confluent Heun function, $q' = q-(1-\gamma)(\epsilon-\delta)$, $\eta' = \eta - (1-\gamma)\epsilon$, $\gamma' = 2-\gamma$. Therefore, the radial wave function $R(x)$ will be given by
\begin{equation}\label{radialwfgeral}
	R(x) = x^{s_{1}}e^{s_{2}x}[C_{1}H_{c}(q,\eta,\gamma,\delta,\epsilon;x) +C_{2}x^{1-\gamma}H_{c}(q',\eta',\gamma',\delta,\epsilon;x)],
\end{equation}
where $x = 1-r/r_{h}$, and it converges in the open unit disk  $-1 < x < 1$, with $\gamma = 2s_{1}+1$, $\delta = 1$, $\epsilon = 2s_{2}$, $q = A$ and $\eta = A + B$. The confluent Heun function can be represented by a power series given by
\begin{equation}
	H_{c}(q,\eta,\gamma,\delta,\epsilon;x) = \sum_{i = 0}^{\infty} \tau_{i}x^{i},
\end{equation}
where the coefficients must obey a three-term recurrence relation \cite{ronveaux, kristensson}, 
\begin{equation}\label{conditionHeuncon}
	A_{i+1}\tau_{i+1} + B_{i}\tau_{i} + D_{i-1}\tau_{i-1} = 0,
\end{equation}
where $ \tau_{-1} = 0$ and $\tau_{0} = 1$, with $A_{i} = i(\gamma + i-1)$, $B_{i} = q - i(i+\gamma+\delta -\epsilon -1)$ and $D_{i} = -\eta - \epsilon i$.

\subsubsection{Bound states and quasispectrum}

In this section, we determine the resonant frequencies, or quasispectrum of energy. To do this, we need to truncate the power series, thereby obtaining the confluent Heun polynomial \cite{kerr-sen}. Therefore, using (\ref{conditionHeuncon}) and imposing the truncation of the series at a given $i = n$, with $n \geq 1$, we obtain the following conditions \cite{fiziev, ishkhan}
\begin{equation}
  D_{n-1} = 0 \hspace{1cm} {\rm and} \hspace{1cm}
	 \Delta_{n+1} = \left|\begin{array}{ccccc}
		B_{0} & A_{1} & 0 & 0 & \ldots\\
		D_{0} & B_{1} & A_{2} & 0 &\ldots\\
		0 & D_{1} & B_{2} & A_{3} & \ldots\\
		\vdots & \vdots & \vdots  & \vdots  & \vdots\\
		0 & 0 & 0 & D_{n-1} & B_{n}
			\end{array}\right| = 0.
\end{equation}

By applying the first condition $D_{n-1} = 0$, we obtain $\eta -(1-n)\epsilon = 0$. Thus, using the values of $\eta$ and $\epsilon$, we obtain
\begin{equation}
	\omega_{n} = \frac{i}{2M}\frac{(1-\alpha)^2}{(2-\alpha)}(n+1),
\end{equation}
thereby obtaining the quasispectrum of energy, whose values are purely imaginary. When $\alpha = 0$, we recover the value of the quasispectrum of the black hole without the presence of quintessence-like anisotropic fluid \cite{bezerra1}. It is interesting to note that in this case we can also obtain the bound state by choosing, for example, $s_{1} = -2iM\omega_{n}/(1-\alpha)^2$ and $s_{2} = - 2M\sqrt{m_{0}^{2}(1-\alpha) - \omega_{n}^2}/(1-\alpha)^2$. Thus, we set $C_{2} = 0$ in (\ref{radialwfgeral}), since $x^{1-\gamma}$ diverges at $x = 0$, and $e^{s_{2}x} \rightarrow 0$ when $x \rightarrow \infty$ for $\omega_{n}^2 < m_{0}^2(1-\alpha)$. Therefore, we have
\begin{equation}
	R_{n}(x) = x^{s_{1}}e^{s_{2}x} \sum_{i = 0}^{n-1}c_{i}x^{i},
\end{equation}
with $s_{1} = (1-\alpha)(n+1)/(2-\alpha).$ Finally, to apply the second truncation condition, we need to define the value of $n$. Considering the ground state as an example, we have $n = 1$, hence
\begin{eqnarray}
	\left|\begin{array}{ccc}
		B_{0} & A_{1} \\
		0 & B_{1}
	\end{array}\right| &=& B_{0}B_{1}\nonumber\\
	&=& q(q - \gamma - \delta + \epsilon
	) = 0,\nonumber
\end{eqnarray}
thus we obtain that one of the possible solutions is $q = 0$. Imposing this condition, we have
\begin{equation}
	\frac{l(l+1)}{1-\alpha} + \frac{2(4-\alpha)}{2-\alpha} = 0,
\end{equation}
and solving for $l$, we obtain
\begin{equation}
	l = -\frac{1}{2} \pm \frac{1}{2}\sqrt{1-8\frac{(4-\alpha)(1-\alpha)}{2-\alpha}},
\end{equation}
Thus, we see that the value of the quantum number $l$ is determined directly by $\alpha$ and may, in general, be complex, just like the quasispectrum. In the limit of $\alpha \rightarrow 0$, we obtain $l = -1/2 \pm i\sqrt{15}/2$.

\subsubsection{Hawking's radiation}

In this section, we determine the Hawking radiation \cite{hawking} through the rate of particles that escape from the black hole across the event horizon. Thus, we first determine the wave function in the region of the event horizon with $r \approx r_{h}$ or $x \approx 0$. Therefore, from (\ref{radialwfgeral}) we have
\begin{equation}
	R(r\approx r_{h}) \approx C_{1} (r-r_{h})^{s_{1}} + C_{2}(r-r_{h})^{-s_{1}},
\end{equation}
where $H_{c}(q,\eta,\gamma,\delta,\epsilon;0) = 1$, $s_{1} = -2iM\omega/(1-\alpha)^2$, and the time-dependent wave function for $r \approx r_{h}$ will be
\begin{equation}
	\psi(r,t) \approx [C_{1}(r-r_{h})^{s_{1}} + C_{2}(r-r_{h})^{-s_{1}}]e^{-i\omega t}.
\end{equation}

From now on, we define $\psi_{in}$ and $\psi_{out}$ as the wave functions describing the modes entering and leaving the event horizon, respectively, with
\begin{eqnarray}
	\left\{\begin{array}{l}
		\psi_{in} = (r-r_{h})^{-is'}e^{-i\omega t}, \hspace{1cm} r<r_{h}\\
		\psi_{out} = (r-r_{h})^{is'}e^{-i\omega t}, \hspace{1cm} r>r_{h}\\
		\end{array}
		\right. ,
\end{eqnarray} 
where $s' = 2M\omega/(1-\alpha)^2$ with $s_{1} = -is'$. Furthermore, in order to investigate the problem in a way that is consistent with the results already obtained for Hawking radiation in the literature, we perform a transformation to the Eddington–Finkelstein coordinate, which is given by
\begin{equation}
	dr' = \frac{r_{h}}{2M}\frac{r}{r-r_{h}}dr \approx \frac{r_{h}^{2}}{2M}\frac{dr}{r-r_{h}},
\end{equation}
where we have applied the condition $r \approx r_{0}$. Thus, by integrating both sides we obtain
\begin{equation}
  r' = \frac{r_{h}^{2}}{2M}\ln(r-r_{h}) \hspace{0.7cm}\rightarrow \hspace{0.7cm} r-r_{h} = \exp\left(\frac{2M}{r_{h}^{2}} r'\right),
\end{equation}
thus, we have 
\begin{eqnarray}
	\left\{\begin{array}{l}
		\psi_{in} =e^{-i\omega(t + r')}, \hspace{1cm} r<r_{h}\\
		\psi_{out} = e^{-i\omega(t - r')}, \hspace{1cm} r>r_{h}\\
	\end{array}
	\right. .
\end{eqnarray} 

Introducing now the variable $v = t + r'$, we finally obtain
\begin{eqnarray}
	\left\{\begin{array}{l}
			\psi_{in} = e^{-i\omega v}, \hspace{1cm} r<r_{h}\\
			\psi_{out} = e^{-i\omega v} (r-r_{h})^{4iM\omega /(1-\alpha)^2}, \hspace{1cm} r>r_{h}\
		\end{array}
		\right. ,
	\end{eqnarray}
and since we want to calculate the rate of particles escaping from the event horizon, we will use only the wave function $\psi_{out}$ in the regions $r>r_{h}$ and $r<r_{h}$. For the region $r<r_{h}$, we need to perform an analytic continuation due to the pole at $r = r_{h}$. Thus, by considering a contour in the lower complex semi-plane, we obtain $r-r_{h} = (r_{h}-r)e^{-i\pi}$, and therefore we have that
\begin{eqnarray}
	\left\{\begin{array}{l}
		\psi_{out}(r<r_{h}) = e^{-i\omega v}(r_{h}-r)^{4iM\omega/(1-\alpha)^2}e^{4\pi M\omega/(1-\alpha)^2} \\
		\psi_{out}(r>r_{h}) = e^{-i\omega v} (r-r_{h})^{4iM\omega/(1-\alpha)^2}\\
	\end{array}
	\right. ,
\end{eqnarray} 
and using the formulation of Sannan \cite{sannan}, the decay rate is
\begin{equation}
	\Gamma = \left|\frac{\psi_{out}(r>r_{h})}{\psi_{out}(r<r_{h})}\right|^2 = \exp\left[-\frac{8\pi M \omega}{(1-\alpha)^2}\right].
\end{equation}

Thus, the Hawking radiation is
\begin{equation}
	N_{H} = \frac{\Gamma}{1-\Gamma} = \left[\exp\left(\frac{8\pi M \omega}{(1-\alpha)^2}\right)-1\right]^{-1},
\end{equation}
and by comparing this result with the Bose–Einstein statistics, which is given by $N_{BE} = [\exp(\omega/k_{B}T)-1]^{-1}$, we obtain that the Hawking temperature, or black hole temperature, is given by
\begin{equation}
	\frac{\omega}{k_{B}T_{H}} = \frac{8\pi M \omega}{(1-\alpha)^2} \hspace{1cm}\rightarrow \hspace{1cm} T_{H} = \frac{(1-\alpha)^2}{8\pi k_{B}M }.
\end{equation}
 
We note that the Hawking temperature obtained here is the same as that derived using other methods, such as the surface gravity approach, for which the Hawking temperature is defined as $T_{H} = f'(r_{h})/(4\pi k_{B})$, where $f(r) = 1-\alpha - 2M/r$. Furthermore, dark energy directly influences both the Hawking radiation and the black hole temperature. In the figures below, we analyze the behavior of the radiation for different values of $\alpha$.

\graphicspath{{figuras/}}

\begin{figure}[!htb]
	\centering
	\includegraphics[scale={0.7}]{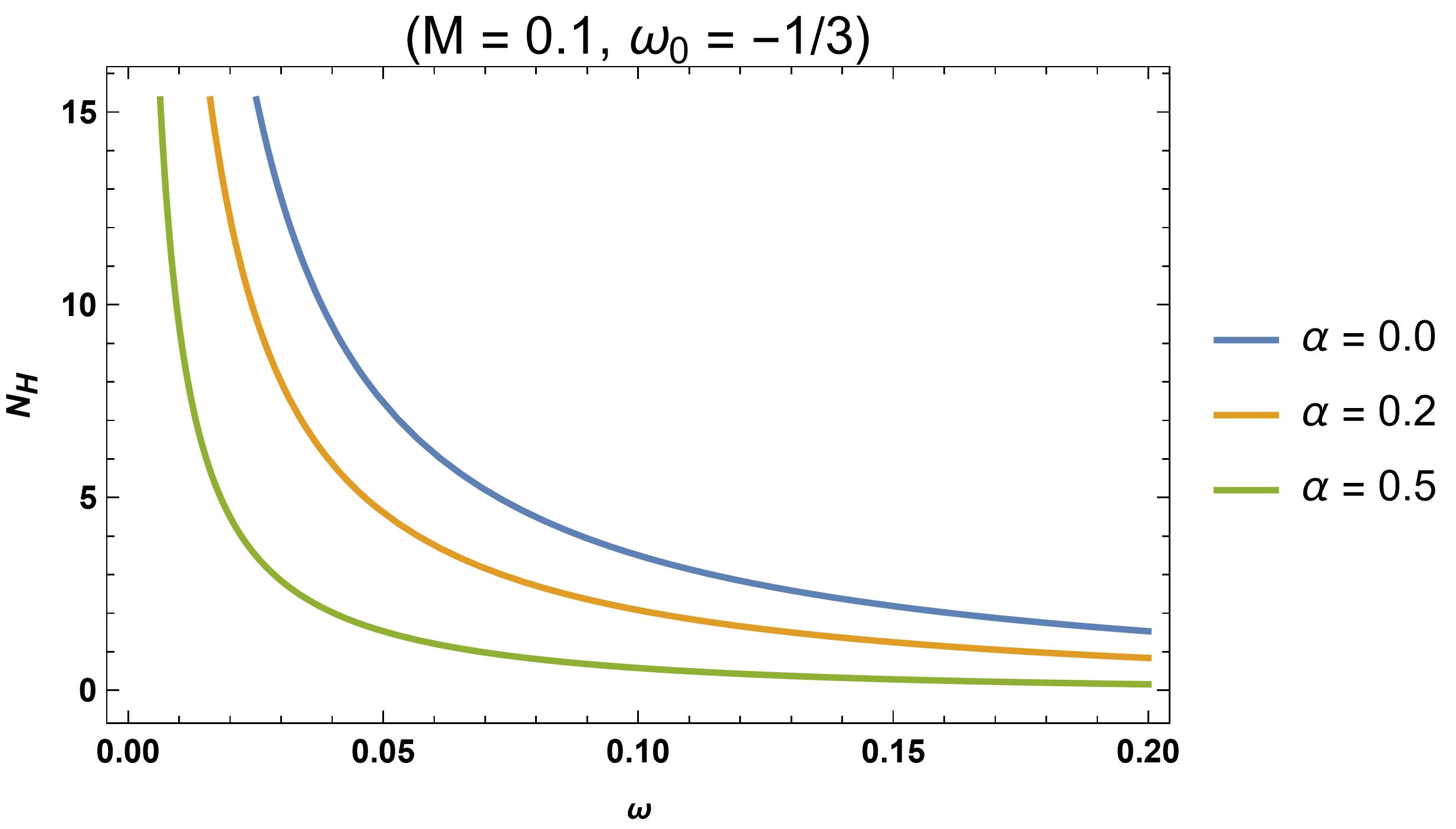}
	\caption{Representation of the behavior of Hawking radiation for $\omega_{0} = -1/3$, in terms of energy $\omega$ for different values of $\alpha$.}
	\label{figuraradiotion1}
\end{figure}

We observe in Fig. (\ref{figuraradiotion1}) that for the Kiselev black hole model with $\omega_{0} = -1/3$, the larger the value of $\alpha$, which is associated with the intensity of the influence of quintessence-like anisotropic fluid, the smaller the Hawking radiation observed in the region outside the event horizon, and consequently the smaller the value of the black hole temperature $T_{H}$. In the limit of maximum quintessence intensity, that is, $\alpha \rightarrow 1$, we obtain $N_{H} \rightarrow 0$ and $T_{H} \rightarrow 0$.

\subsection{Second case: $\omega_{0}  = -2/3$}

In this second and final case, we consider a Kiselev black hole model with $\omega_{0} = -2/3$. Thus, using (\ref{radialequationgeral}), we have
\begin{eqnarray}\label{radialequation2}
	\left[\frac{d^2}{d r^2} + \left(\frac{2}{r} - \frac{2M - \alpha r^2}{\alpha r(r-r_{h})(r-r_{1})}\right)\frac{d}{d r} + \frac{l(l+1)}{\alpha r(r-r_{h})(r-r_{1})} + \frac{m_{0}^{2} r}{\alpha (r-r_{h})(r-r_{1})} \nonumber \right. \\\\ \left.  + \frac{\omega^2 r^2}{\alpha^2 (r-r_{h})^2(r-r_{1})^2}\right]R(r) = 0,\nonumber
\end{eqnarray}
where the physical event horizon is $r_{h} = (1-\sqrt{1-8\alpha M})/(2\alpha)$ and $r_{1} = (1+\sqrt{1-8\alpha M})/(2\alpha)$. Thus, to solve the differential equation above, we again perform the substitution $x = 1 - r/r_{h}$, thus we obtain

\begin{eqnarray}\label{radialequation2x}
	\left\{\frac{d^2}{d x^2} + \left(\frac{1}{x} + \frac{1}{x-1} + \frac{1}{x-a}\right)\frac{d}{d x} + \left[\frac{l(l+1)}{\alpha r_{h}a} + \frac{ m_{0}^{2} r_{h}}{\alpha a} - \frac{2\omega^2}{\alpha^2 a^2}\left(1-\frac{1}{a}\right)\right]\frac{1}{x} + \frac{\omega^2}{\alpha^2 a^2}\frac{1}{x^2} \nonumber \right. \\\nonumber\\ \left. -\frac{l(l+1)}{\alpha r_{h}a(a-1)}\frac{1}{x-1} + \left[\frac{l(l+1)}{\alpha r_{h}a(a-1)} + \left(\frac{m_{0}^{2}r_{h}}{\alpha}+ \frac{2 \omega^2}{\alpha^2 a^2}\right)\left(1-\frac{1}{a}\right) \right]\frac{1}{x-a} \right. \\\nonumber\\ \left. \frac{\omega^2}{\alpha^2}\left(1-\frac{1}{a}\right)^2 \frac{1}{(x-a)^2} \right\}R(x)  = 0,\nonumber
\end{eqnarray}
where $a = 1 - r_{1}/r_{h}$. This time, the solution of this differential equation can be obtained in terms of the local Heun function via the ansatz $R(x) = x^{s_{1}}(x-a)^{s_{2}} f(x)$. eliminate the inverse quadratic terms $1/x^2$ and $1/(x-a)^{2}$ from the equation, and thus we obtain
\begin{eqnarray}\label{radialequation2f}
	\left\{\frac{d^2}{d x^2} + \left(\frac{2s_{1}+1}{x} + \frac{1}{x-1} + \frac{2s_{2}+1}{x-a}\right)\frac{d}{d x} + [(\xi_1 + \xi_2 + \xi_3)x^2 \right.\nonumber \\\\ \left. - (a\xi_1 + (a+1)\xi_2 + \xi_3)x + a\xi_2]\frac{1}{x(x-1)(x-a)}\right\}f(x)  = 0,\nonumber
\end{eqnarray}
where 
\begin{equation}
\left.\begin{array}{l}
\displaystyle \xi_{1} =  s_{1} - \frac{s_{2}}{a-1} - \frac{l(l+1)}{\alpha r_{h}(a-1)}\\\\
\displaystyle \xi_{2} = \frac{l(l+1)}{\alpha r_{h}a} + \frac{m_{0}^{2}r_{h}}{\alpha a} - \frac{2\omega^2}{\alpha^2 a^2}\left(1-\frac{1}{a}\right) - \left(1+\frac{1}{a}\right)s_{1} -\frac{s_{2}}{a} - \frac{2s_{1}s_{2}}{a}\\\\
\displaystyle \xi_{3} = \frac{l(l+1)}{\alpha r_{h}a(a-1)} + \left(\frac{m_{0}^{2}r_{h}}{\alpha}+ \frac{2\omega^2}{\alpha^2 a^2}\right)\left(1-\frac{1}{a}\right)  +  \frac{s_{1}}{a} + \left(\frac{1}{a} + \frac{1}{a-1}\right)s_{2} + \frac{2s_{1}s_{2}}{a} 
\end{array}
\right.,
\end{equation}
with $s_{1} =  \pm i\omega/(\alpha a)$ and $s_{2} = \pm i\omega(a-1)/(\alpha a)$. The equation (\ref{radialequation2f}) is analogous to the local Heun equation, which has three finite singularities at $z = (0,1,a)$ and one at infinity $z \rightarrow \infty$, and in a generalized form is given by \cite{schafker, kerr-de-sitter}
\begin{equation}\label{heunequation}
	y''(z) + \left(\frac{1-\mu_0}{z} + \frac{1-\mu_1 }{z-1} + \frac{1-\mu_2}{z-a_{H}} + \alpha'\right)y'(z) + \frac{(\beta_2 z^2 + \beta_1 z + \beta_0)}{z(z-1)(z-a_H)}y(z) = 0,
\end{equation}
whose solution is 
\begin{equation}
	y(z) = c_{1} \eta(z,\mu_0,\mu_1,\mu_2,\alpha',\lambda;a_H) + c_{2} z^{\mu_0}\eta(z,-\mu_0,\mu_1,\mu_2,\alpha',\lambda;a_H),
\end{equation}
where the generalized Heun function is defined by 
\begin{equation}
	\eta(z,\mu_0,\mu_1,\mu_2,\alpha',\lambda;a_H) = \sum_{k=0}^{\infty} \frac{\tau_{k}(\mu,\alpha',\lambda;a_H)}{\Gamma(1+k -\mu_0)\Gamma(1+k)}x^{k},
\end{equation}
where $\Gamma(z)$ is the Euler gamma function, and the parameters $\mu$ and $\lambda$ denote the sets of values $(\mu_0,\mu_1,\mu_2)$ and $(\lambda_0,\lambda_1,\lambda_2)$, respectively. In our case, we identify $\alpha' = 0$, $\mu_0 = -2s_1$, $\mu_1 = 0$, $\mu_2 = -2s_2$, $\beta_2 = \xi_1 + \xi_2 + \xi_3$, $\beta_1 = -(a\xi_1 + (a+1)\xi_2 + \xi_3)$, $\beta_0 = a\xi_2$ and $a_{H} = a$. The parameters $\lambda$ must satisfy the relation
\begin{equation}
	\frac{\beta_2 z^2 + \beta_1 z + \beta_0}{z(z-1)(z-a_H)} = \sum_{\substack{\sigma,\rho =0 \\ \sigma\neq \rho}}^{2}\frac{1}{2} \left(\frac{1-\mu_{\sigma}}{z-z_{\sigma}}\right)\left(\frac{1-\mu_{\rho}}{z-z_{\rho}}\right) + \sum_{k = 0}^{2} \frac{\alpha'(1-\mu_k)/2 + \lambda_k}{z-z_k},
\end{equation} 
with $z_{0} = 0$, $z_{1} = 1$ and $z_2 = a$. The coefficients $\tau_{k}$ are obtained through the four-term recurrence relation given by
\begin{equation}\label{recorrence}
	A_{k+1}\tau_{k+1} + B_{k}\tau_{k} + C_{k-1}\tau_{k-1} + D_{k-2}\tau_{k-2} = 0,
\end{equation}
where $\tau_{-2} = \tau_{-1} = 0$ and $\tau_{0} = 1$ with
\begin{eqnarray}
	\left. \begin{array}{l}
		\displaystyle	A_{k} = 1 \\
		\displaystyle	B_{k} = \frac{\beta_0}{a_H} + \alpha'(k+1) -(k+1)(2+k-\mu_0-\mu_1)-\frac{(k+1)}{a_H} (2+k-\mu_0-\mu_2)\\\\
		\displaystyle C_{k} = (k+2)(k+2-\mu_0)\left[\frac{(k+1)}{a_H}(3+k-\mu_0-\mu_1-\mu_2) -\left(1+\frac{1}{a_H}\right)(k+1)\alpha'+\frac{\beta_1}{a_H}\right]\\\\\
		\displaystyle D_{k} = \frac{(k+2)(k+3)}{a_H}(k+2-\mu_0)(k+3-\mu_0)(\alpha'(k+1)+\beta_2)
	\end{array}\right..
\end{eqnarray}

Applying (\ref{recorrence}) for $k=0,1,\ldots,\infty$, we obtain
\begin{eqnarray}\label{relacaomatricial}
	\left(\begin{array}{cccccc}
		B_{0} & A_1 & 0 & 0 & 0 & \ldots \\
		C_0 & B_1 & A_2 & 0 & 0 & \ldots \\
		D_{0} & C_1 & B_2 & A_3 & 0 & \ldots\\
		0 &  D_{1} & C_2 & B_3 & A_4  & \ldots\\
		\vdots & \vdots & \vdots & \vdots & \vdots
	\end{array}\right) \left(\begin{array}{c}
		\tau_0\\
		\tau_1\\
		\tau_2\\
		\tau_4\\
		\vdots
	\end{array}
	\right) = \left(\begin{array}{c}
		0\\
		0\\
		0\\
		0\\
		\vdots
	\end{array}
	\right).
\end{eqnarray}

Thus, the radial wave function will be given by 
\begin{eqnarray}\label{radialwfgeral2}
	R(x) &=& x^{s_{1}}(x-a)^{s_{2}}[c_{1} \eta(x,-2s_1,0,-2s_2,0,\lambda;a) + c_{2} x^{1-2s_1}\times\nonumber \\ && \eta(x,2s_1,0,-2s_2,0,\lambda;a)],
\end{eqnarray}
where $0 \leq x < \infty $.

\subsubsection{Quasibound states and quasispectrum}

In this section, we determine the quasispectrum of energy. To do this, we again need to truncate the power series, thereby obtaining the generalized local Heun polynomial. It is worth emphasizing that bound states cannot be obtained throughout the entire domain of $x$, since the radial wave function given by (\ref{radialwfgeral2}) remains regular but does not approach zero at either $x=0$ or $x\rightarrow\infty$. Consequently, by requiring that $R(x)$ remain finite at both $x=0$ and $x\rightarrow\infty$, we obtain the quasibound states together with the corresponding resonant frequencies, or, more generally, the quasispectrum. To satisfy these conditions, the infinite series that defines the function $\eta$ must terminate, ensuring that it does not diverge as $x\rightarrow\infty$. Therefore, by truncating the series at a given $i = n$, with $n \geq 2$, we obtain $(n+1)$ homogeneous linear equations and the following conditions analogous to those obtained in \cite{schafker, kerr-de-sitter},
\begin{equation}
	D_{n-2} = 0 \hspace{1cm} {\rm and} \hspace{1cm}
	\Delta_{n+1} = \left|\begin{array}{ccccc}
		B_{0} & A_{1} & 0 & 0 & \ldots\\
		C_{0} & B_{1} & A_{2} & 0 &\ldots\\
		D_{0} & C_{1} & B_{2} & A_{3} & \ldots\\
		\vdots & \vdots & \vdots  & \vdots  & \vdots\\
		0 & 0 & 0 & D_{n-2} & C_{n-1}
	\end{array}\right| = 0.
\end{equation}

By applying the first condition $D_{n-2} = 0$, we obtain $\beta_2 +(n-1)\alpha' = 0$, and using the value $\alpha' = 0$ we have $\beta_2 = 0$. Hence,
\begin{equation}
	\beta_2 = \xi_{1} + \xi_{2} + \xi_{3} = \frac{m_0^{2} r_{s}}{\alpha} = 0 \hspace{1cm}\rightarrow \hspace{1cm} m_{0} = 0,
\end{equation}
thus we obtain that one of the conditions required to truncate the series is that the particle mass must be zero.  Furthermore, it is important to note that $\beta_2 = 0$, regardless of the value of $k$. Consequently, the recurrence relation in (\ref{recorrence}) reduces to a three-term recurrence relation, and the equation in (\ref{heunequation}) becomes of the local Heun type. Therefore, to truncate the series, we now require that $\tau_{n-1}\neq0$ and $C_{n-1}=0$, which ensures that $\tau_{n+1}=\tau_n=0$ for $n \geq 1$. Hence, by imposing $C_{n-1}=0$ and using the expression for $\beta_1$, we obtain
\begin{equation}\label{quasispectrum}
	\omega_n = -i\frac{\alpha a (a-2)(n+1)}{4(a-1)} - \frac{i\alpha a}{4(a-1)}\sqrt{4+a[a(n+1)^2-4]},
\end{equation}
where we use $s_{1} = -i\omega/(\alpha a)$, $s_{2} = i(a-1)\omega/(\alpha a)$ and $a = 1-r_{1}/r_{h}$. In the limit $\alpha \rightarrow 0$, we once again recover the quasispectrum of the Schwarzschild black hole, given by $\omega = i(n+1)/4M$.  Using the second condition given by $\Delta_{n+1} = 0$, we can obtain the value of a physical parameter such as, for example, the angular momentum $l$ for each specific value of $n$. In other words, these values will be quantized and will assume complex values due to parameters with complex values such as $s_{1}$ and $s_{2}$. Consequently, the values of $\omega$ will also be complex and will take on different values for each value of $n$, and are referred to as the quasispectrum. For the ground state, for example, we have $n = 1$; therefore, by imposing $\Delta_{2} = 0$, we obtain
\begin{equation}
	\left|\begin{array}{ccc}
		B_{0} & A_{1}\\
		0 & B_{1} 
	\end{array}\right| = B_{0}B_{1}  = 0,
\end{equation}
thus, we have 
\begin{eqnarray}
	\left[\xi_2  -2\left(1-\frac{i\omega}{\alpha a}\right) -\frac{2}{a}\left(1-\frac{2i\omega}{\alpha a}+ \frac{i\omega}{\alpha }\right)\right]\left[\xi_2  -2\left(3-\frac{2i\omega}{\alpha a}\right)\nonumber \right. \\\\ \left. -\frac{2}{a}\left(3-\frac{4i\omega}{\alpha a} + \frac{2i\omega}{\alpha }\right)\right] = 0,\nonumber
\end{eqnarray}
and we can obtain, for example, the values of the complex angular momentum, as in the previous case.
 The radial wave function will be 
\begin{eqnarray}\label{Rfinal}
	R(x) &=& x^{s_1}(x-a)^{s_2}[c_{1} H_{l}(a,\beta_0,\beta_1,\mu_0,\mu_1,\mu_2,\lambda;x) + c_{2}x^{-2s_1}\times \nonumber \\ &&  H_{l}(a,\beta_0,\beta_1,-\mu_0,\mu_1,\mu_2,\lambda;x)],
\end{eqnarray}
with $s_{1} = -i\omega/(\alpha a)$, $s_{2} = i\omega (a-1)/(\alpha a)$, and the local Heun function $H_l$ satisfies Eq.~(\ref{heunequation}) when $\alpha'=\beta_2=0$. Finally, we must verify that the wave function remains finite as $x\rightarrow\infty$. To this end, we consider the Heun equation (\ref{heunequation}) in the neighborhood of the irregular singular point at infinity, where two independent solutions exist. Expanding the local Heun function in this region yields the following asymptotic series \cite{ronveaux}
\begin{equation}\label{Heunassintotico}
	H_{l}(a,\beta_0,\beta_1,\mu_0,\mu_1,\mu_2,\lambda;x)] \approx c_{1}x^{-\rho_+} + c_{2}x^{-\rho_{-}}
\end{equation}
where we retain only the leading terms of the asymptotic power series, with $\rho_{\pm}=s_1+s_2+1\pm i\chi$ and $\chi=-\sqrt{\beta_1-\left(s_1+s_2+1\right)^2}$. Therefore, from equations (\ref{Rfinal}) and (\ref{Heunassintotico}), the radial wave function far from the black hole, i.e., in the limit $|x|\rightarrow\infty$, is given by
\begin{equation}\label{radialfunction1}
	R(x) \approx \frac{1}{x}[c_{1}x^{-i\chi} + c_{2}x^{i\chi}] =  \frac{1}{x}[c_{1} e^{-i\chi \ln(x)} + c_{2} e^{i\chi \ln(x)}].
\end{equation}

Thus, the radial wave function can be written as
\begin{equation}
	R(x) \approx c_{\lambda} \frac{1}{x}\sin[\chi {\rm ln}(x) + \sigma_{\lambda}(\omega)],
\end{equation}
where $\sigma_{\lambda}(\omega)$ is the phase shift. We therefore verify that the wave function remains finite in the limit $x\rightarrow\infty$, as required. Moreover, these solutions can be used to investigate the scattering of both massive and massless scalar fields. The parameter $\sigma_{\lambda}$ can be determined approximately following the procedure developed by Abramov \textit{et al.} \cite{abramov}.

\subsubsection{Hawking's radiation}

In this section, we calculate the Hawking radiation through the rate of particles that escape from the black hole across the event horizon. To do this, we first determine the wave function in the region of the event horizon with $r \approx r_{h}$ or $x \approx 0$. Thus, from (\ref{radialwfgeral2}) we have
\begin{equation}
	R(r\approx r_{h}) \approx C_{1} (r-r_{h})^{s_{1}} + C_{2}(r-r_{h})^{-s_{1}},
\end{equation}
where we use $H_{l}(a,q,\gamma,\delta,\epsilon,\mu,\eta,\beta;0)=1$, and the time-dependent wave function for $r \approx r_{h}$ will be
\begin{equation}
	\psi(r,t) \approx [C_{1}(r-r_{h})^{s_{1}} + C_{2}(r-r_{h})^{-s_{1}}]e^{-i\omega t},
\end{equation}
where $s_{1} = -i\omega/(\alpha a)$. As in the previous case, we define $\psi_{in}$ and $\psi_{out}$ as the wave functions describing the modes entering and leaving the event horizon, respectively, with
\begin{eqnarray}
	\left\{\begin{array}{l}
		\psi_{in} = (r-r_{h})^{is'}e^{-i\omega t}, \hspace{1cm} r<r_{h}\\
		\psi_{out} = (r-r_{h})^{-is'}e^{-i\omega t}, \hspace{1cm} r>r_{h}\\
	\end{array}
	\right. ,
\end{eqnarray} 
with $s' = \omega/(\alpha a)$, where we use $s_{1}  = -i s'$ and $\alpha a < 0$. Now, we perform the transformation to the Eddington–Finkelstein coordinate, which is given by
\begin{equation}
	dr' = -\frac{r}{\alpha(r-r_{h})(r-r_{1})}dr \approx -\frac{dr}{a\alpha(r-r_{h})},
\end{equation}
where we have applied the condition $r \approx r_{h}$. Thus, by integrating both sides we obtain
\begin{equation}
	r' = -\frac{1}{a\alpha}\ln(r-r_{h}) \hspace{0.7cm}\rightarrow \hspace{0.7cm} r-r_{h} = e^{-a\alpha r'},
\end{equation}
thus, we have  
\begin{eqnarray}
	\left\{\begin{array}{l}
		\psi_{in} = e^{-i\omega(t + r')}, \hspace{1cm} r<r_{h}\\
		\psi_{out} =  e^{-i\omega(t - r')}, \hspace{1cm} r>r_{h}\\
	\end{array}
	\right. .
\end{eqnarray} 

Introducing now the variable $v = t + r'$, we finally obtain
\begin{eqnarray}
	\left\{\begin{array}{l}
		\psi_{in} = e^{-i\omega v}, \hspace{1cm} r<r_{h}\\
		\psi_{out} = e^{-i\omega v} (r-r_{h})^{-2is'}, \hspace{1cm} r>r_{h}\\
	\end{array}
	\right. ,
\end{eqnarray} 
and since we want to calculate the rate of particles escaping from the event horizon, we will use only the wave function $\psi_{out}$ in the regions $r>r_{h}$ and $r<r_{h}$. For the region $r<r_{h}$, we need to perform an analytic continuation due to the pole at $r = r_{h}$. Thus, by considering a contour in the lower complex semi-plane, we obtain $r-r_{h} = (r_{h}-r)e^{-i\pi}$, and therefore we have that
\begin{eqnarray}
	\left\{\begin{array}{l}
		\psi_{out}(r<r_{h}) = e^{-i\omega v}(r_{h}-r)^{-2is'}e^{-2\pi s'} \\
		\psi_{out}(r>r_{h}) = e^{-i\omega v} (r-r_{h})^{-2is'}\\
	\end{array}
	\right. ,
\end{eqnarray} 
and using the formulation of Sannan once again, the decay rate will be given by
\begin{equation}
	\Gamma = \left|\frac{\psi_{out}(r>r_{h})}{\psi_{out}(r<r_{h})}\right|^2 = \exp\left(\frac{4\pi \omega}{\alpha a}  \right),
\end{equation}
where $\alpha a = \alpha (r_{h}-r_{1})/r_{h} < 0$. With that, the Hawking radiation will be
\begin{equation}
	N_{H} = \frac{\Gamma}{1-\Gamma} = \left[\exp\left( -\frac{4\pi \omega}{\alpha a} \right)-1\right]^{-1}.
\end{equation}

By comparing this result with the Bose–Einstein statistics, which is given by $N_{BE} = [\exp(\omega/k_{B}T)-1]^{-1}$, we obtain that the Hawking temperature is given by $\omega/(k_{B}T_{H}) = -4\pi \omega/(\alpha a)$, and solving it, we obtain that the Hawking temperature is
\begin{equation}
	T_{H} = -\frac{\alpha a}{4\pi k_{B}} = \frac{\alpha(r_1 - r_h)}{4\pi k_{B}r_{h}},
\end{equation}
we observe that, in this case, the Hawking temperature no depends on the energy of the particle that escapes from the event horizon. We also note that, once again, the Hawking temperature obtained here coincides with that derived using other methods, such as the surface gravity approach, for which the Hawking temperature is defined as $T_{H} = f'(r_{h})/(4\pi k_{B}) = \alpha(r_{1}-r_{h})/(4\pi k_{B} r_{h})$, where $f(r) = 1-\alpha r - 2M/r$. In the limit $\alpha \rightarrow 0$, we recover the value $T_{H} = 1/(8\pi k_{B}M)$, which is the Hawking temperature for a black hole without the presence of quintessence \cite{bezerra1}. We therefore observe that both dark energy and the particle's energy directly affect the Hawking radiation, whereas the black hole temperature is influenced solely by the quintessence field. In the figure below, we analyze the behavior of the radiation for different values of $\alpha$ and $\omega$.

\graphicspath{{figuras/}}

\begin{figure}[!htb]
	\centering
	\includegraphics[scale={0.7}]{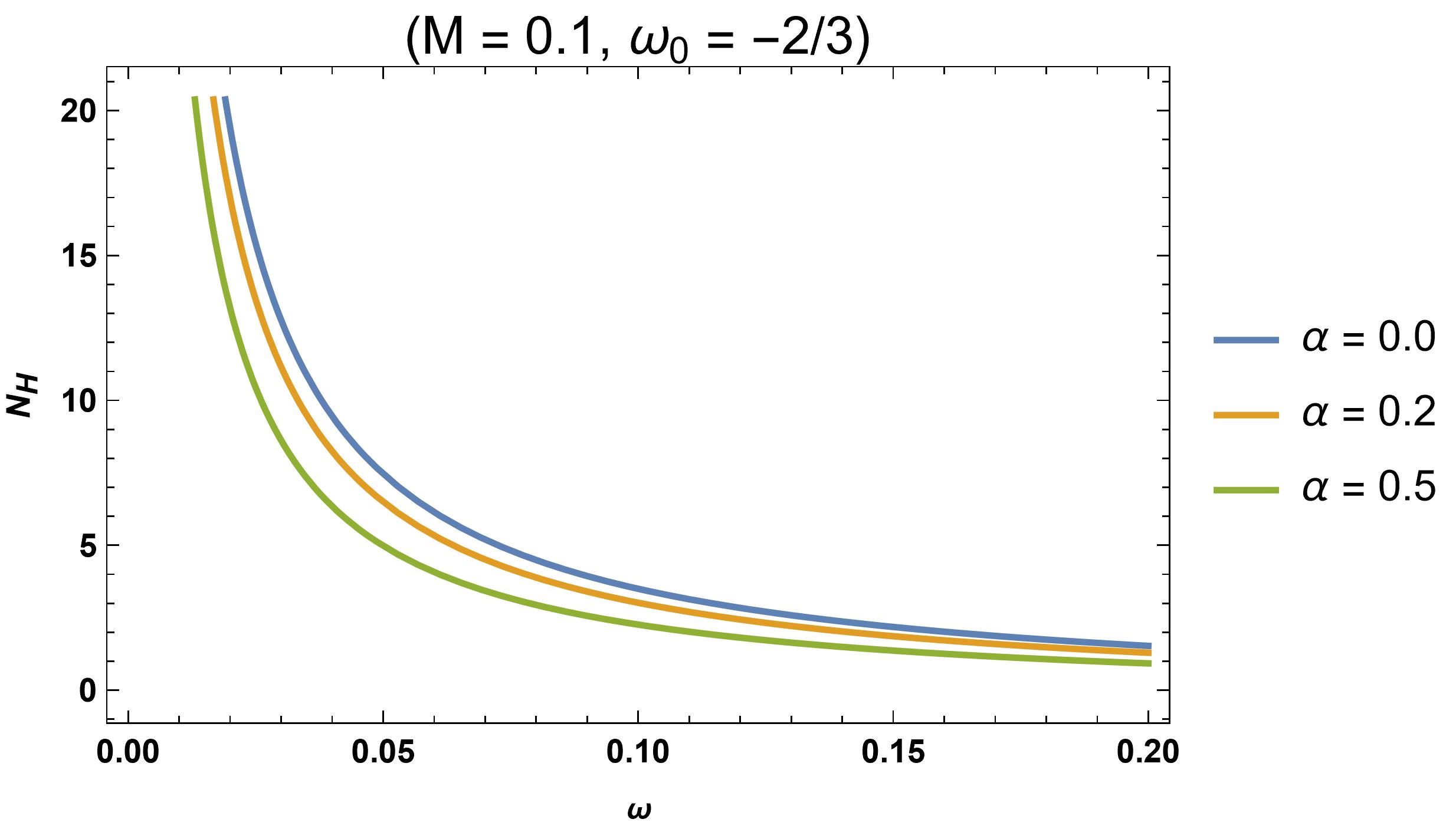}
	\caption{Representation of the behavior of Hawking radiation for $\omega_{0} = -2/3$, in terms of energy $\omega$ for different values of $\alpha$.}
	\label{figuraradiotion2}
\end{figure}

We observe in Fig. (\ref{figuraradiotion2}) that for the quintessence model with $\omega_{0} = -2/3$, the stronger the influence of dark energy, that is, the larger the value of $\alpha$, the smaller the Hawking radiation observed in the region outside the event horizon for values of $\omega>0$.

\section{Conclusion}

We investigate exactly the dynamics of a relativistic spinless particle, through the radial wave function, influenced by a static and uncharged black hole surrounded by a quintessence-like anisotropic fluid or Kiselev black hole. Considering two examples of Kiselev black hole models, we find that the radial equation is given by the confluent Heun equation and the generalized local Heun equation, as is frequently obtained in this type of analysis involving black holes in various contexts \cite{hortacsu,bezerra1,vieira-2015,vieira-class-quantum-grav,Senjaya1,Senjaya2,Senjaya3,Senjaya4}. After obtaining the wave function for the two particular Kiselev black hole models, we determine the bound states for the first case in Section (3.1) and, consequently, the quasispectrum of energy. For the second case, in Section (3.2), we also determine the quasibound states and obtain the quasispectrum of energy $\omega$ for the massless case.

In both cases, we determine the Hawking radiation and temperature, and we verify in Figs. (\ref{figuraradiotion1}) and (\ref{figuraradiotion2}) that the stronger the influence of quintessence, that is, the larger the value of the parameter $\alpha$, the smaller the radiation observed outside the event horizon. This result is interesting, as it may provide further evidence for the existence of dark energy in the universe, since, in fact, very little or no matter is observed outside the event horizon of a black hole, as suggested by several theoretical studies in the literature.

Another interesting feature is that, even in the presence of a quintessence-like anisotropic fluid, the Hawking temperature is independent of the energy of the emitted particles. Furthermore, the values of the Hawking temperature obtained for the cases with $\omega_{0} = -1/3$ and $\omega_{0} = -2/3$ coincide with those calculated using other methods \cite{tharanath,hamil}, such as the surface gravity approach, in which the Hawking temperature is given by $T_{H} = f'(r_{h})/(4\pi k_{B})$, where $f(r) = 1-\alpha r^{-1-3\omega_{0}} - 2M/r$.

Finally, in the limit $\alpha \rightarrow 0$, we recover in both cases the Hawking temperature $T_{H} = 1/(8\pi k_{B} M)$ corresponding to a black hole without the presence of dark energy \cite{bezerra1}.

\section*{Acknowledgments}

AGMS gratefully acknowledges CNPq (grant number 309052/2023-8) for partial financial support. This study was funded by FAPERJ - Fundação Carlos Chagas Filho de Amparo à Pesquisa do Estado do Rio de Janeiro, Processess SEI 26/200.337/2024 and SEI 260003/021954/2025.

\section*{Data Availability Statement}

No Data associated in the manuscript.


\section*{References}

\begin{thebibliography}{10}
	
	
		\bibitem{kerr}
	R. P. Kerr,
	\newblock{Phys. Rev. Lett.} \textbf{11}, 5 (1963).
	
	\bibitem{gravitation}
	C. W. Misner, K. S. Thorne and J. A. Wheeler,
	\newblock {\em Gravitation}.
	\newblock Princeton University Press, (2017).	
	
	\bibitem{riess}
	 A. G. Riess et al., 
	\newblock{Astron. J.} {\bf 116}, 1009 (1998).
	
	\bibitem{perlmutter}
	 S. Perlmutter et al., 
	\newblock{ Astrophys. J.} {\bf 517}, 565 (1999).
	
	\bibitem{planck}
	Planck Collaboration,  Planck 2018 results. VI. Cosmological parameters,
	\newblock{Astron. Astrophys.} {\bf 641}, A6 (2020).

	\bibitem{ratra}
    B. Ratra and P. J. E. Peebles,
	\newblock{Phys. Rev. D} {\bf 37}, 3406 (1988).
	
	\bibitem{caldwell}
	R. R. Caldwell, R. Dave, and P. J. Steinhardt,
	\newblock{Phys. Rev. Lett.} {\bf 80}, 1582 (1998).
	
	
	\bibitem{tsujikawa}
    S. Tsujikawa, 
	\newblock{Class. Quantum Grav.} {\bf 30}, 214003 (2013).
	
	\bibitem{kiselev}
	V. V. Kiselev,
	\newblock{Class. Quantum Grav.} \textbf{20}, 1187 (2003).
	
		\bibitem{visser}
	M. Visser,
	\newblock{Class. Quantum Grav.} \textbf{37}, 045001 (2020).
	
	\bibitem{zhou}
	X. Zhou, J. Chen, Y. Wang, and Y. Wang,
	\newblock{Int. J. Mod. Phys. D} \textbf{28}, 1950023 (2019).
	
		\bibitem{liu}
    Y. Liang, X. Yang, and Y. Liu,
	\newblock{Eur. Phys. J. C} \textbf{80}, 808 (2020).
	
		\bibitem{ma}
	M. Ma,
	\newblock{Phys. Lett. B} \textbf{807}, 135535 (2020).
	
		\bibitem{tsupko}
	O. Yu. Tsupko, 
	\newblock{Phys. Lett. D} \textbf{95}, 104058 (2017).
	
		\bibitem{amarilla}
	L. Amarilla and E. F. Eiroa,
	\newblock{Phys. Lett. D} \textbf{85}, 064019 (2012).
	
		\bibitem{wei}
	S.-W. Wei and Y.-X. Liu, 
	\newblock{JCAP} \textbf{11}, 063 (2013).
	
	\bibitem{konoplya0}
	R. A. Konoplya and A. Zhidenko, 
	\newblock{ Rev. Mod. Phys.} \textbf{83}, 793 (2011).
	
	\bibitem{berti}
	E. Berti et al., 
	\newblock{ Class. Quantum Grav.} \textbf{32}, 243001 (2015).
	

\bibitem{hatsuda}
Y. Hatsuda,
\newblock { Class. Quantum Grav.} {\bf 38}, 025015 (2021).


\bibitem{suzuki}
H. Suzuki, E. Takasugi, H. Umetsu,
\newblock { Prog. Theor. Phys.,} {\bf 100}, 491 (1998).

\bibitem{vieira}
H. S. Vieira, V. B. Bezerra and C. R. Muniz, 
\newblock {\em Ann. Phys.} {\bf 350}, 14 (2014).

\bibitem{vieira-2015}
H. S. Vieira, V. B. Bezerra, G. V. Silva, Ann. Phys. {\bf 362}, 576 (2015).

\bibitem{vieira-class-quantum-grav}
V. B. Bezerra, H. S. Vieira, A. A. Costa, Class. Quantum Grav. {\bf 31}, 045003 (2014).

\bibitem{bezerra1}
H. S. Vieira and V. B. Bezerra,
\newblock{ Ann. Phys.} {\bf 373}, 28 (2016).

\bibitem{hortacsu}
M. Horta\c csu, Adv. High Energy Phys. {\bf 2018}, 8621573 (2018).

\bibitem{Senjaya1}
D. Senjaya,
Eur. Phys. J. C  {\bf 84}, 57 (2024).


\bibitem{Senjaya2}
D. Senjaya,
Eur. Phys. J. C  {\bf 84}, 229 (2024).


\bibitem{Senjaya3}
D. Senjaya,
Eur. Phys. J. C  {\bf 84}, 388 (2024).


\bibitem{Senjaya4}
D. Senjaya,
Eur. Phys. J. C  {\bf 84}, 424 (2024).


\bibitem{Senjaya5}
D. Senjaya,
Eur. Phys. J. C  {\bf 84}, 607 (2024).


\bibitem{Senjaya6}
D. Senjaya,
Phys. Lett. B  {\bf 849}, 138414 (2024).


\bibitem{Senjaya7}
D. Senjaya,
Journal of High Energy Astrophysics,  {\bf 40}, 49 (2023).


\bibitem{Senjaya8}
D. Senjaya,
Phys. Lett. B  {\bf 848}, 138373 (2024).


\bibitem{Senjaya9}
D. Senjaya,
Journal of High Energy Astrophysics,  {\bf 42}, 197 (2024).

\bibitem{kokkotas}
H. S. Vieira, K. Destounis and K. D. Kokkotas,
Phys. Rev. D  {\bf 107}, 104038 (2023).

\bibitem{bekenstein}
J. D. Bekenstein,
\newblock{Phys. Rev. D} {\bf 7}, 2333 (1973).

\bibitem{hawking}
S. W. Hawking,
\newblock{ Comm. Math. Phys.} {\bf 43}, 199 (1975).

\bibitem{saleh}
M. Saleh et al.,
\newblock{Eur. Phys. J. C} {\bf 79}, 1019 (2019).

\bibitem{ghoshhawking}
S. G. Ghosh and R. Kumar,
\newblock{  Class. Quantum Grav.} {\bf 35}, 095008 (2018).

	\bibitem{chen}
S. Chen and J. Jing,
\newblock{Class. Quantum Grav.} \textbf{22}, 4651 (2005).

\bibitem{jing}
J. Jing,
\newblock{Phys. Rev. D} \textbf{72}, 027501 (2005).

\bibitem{ghosh}
S. G. Ghosh,
\newblock{Eur. Phys. J. C} \textbf{76}, 222 (2016).

\bibitem{fernando}
S. Fernando,
\newblock{Gen. Relativ. Gravit.} \textbf{44}, 1857 (2012).


\bibitem{arfken}
G. B. Arfken, H. J. Weber, F. E. Harris, {\it Mathematical Methods for Physicists}, $7^{th}$ edition, Academic Press (2013).


\bibitem{ronveaux}
A. Ronveaux,
\newblock {\em  Heun's Differential Equation}.
\newblock Oxford Univ. Press, (1995).

\bibitem{kristensson}
G. Kristensson, {\it Second Order Differential Equations}, Springer (2010). See the excellent chapter on Heun's equation.

\bibitem{kerr-sen}
A. G. M. Schmidt, M. E. Pereira, J. Math. Phys. {\bf 65}, 122501 (2024). In the appendix of this paper there is a recipe to calculate confluent Heun's polynomial. 

\bibitem{fiziev}
P. P. Fiziev,  Class. Quantum Grav. {\bf 27}, 135001 (2010).

\bibitem{ishkhan}
A. M. Ishkhanyan and D.Yu.Melikdzhanian,
\newblock{ J. Math. Anal. Appl.} {\bf 499}, 125037 (2021).


\bibitem{sannan}
S. Sannan,
\newblock{Gen. Relativ. Gravit.} {\bf 20}, 239 (1988).


\bibitem{schafker}
R. Schäfke and D. Schmidt, 
\newblock{SIAM J. Math. Anal.}  {\bf 11}, 848 (1980).

\bibitem{kerr-de-sitter}
A. G. M. Schmidt, M. E. Pereira, Ann. Phys. {\bf 458}, 169465 (2023).

\bibitem{abramov}
D. I. Abramov, A. Y. Kazakov, L. I. Ponomarev, S. Y. Slavyanov and L. N. Somov, 
\newblock{J. Phys. B}  {\bf 12}, 1761 (1979).

\bibitem{tharanath}
R. Tharanath and V. C. Kuriakose, 
\newblock{Mod. Phys. Lett. A}  {\bf 28}, 1350003 (2013).


\bibitem{hamil}
B. C. Lütfüo\v{g}lu, B. Hamil and L. Dahbi, 
\newblock{Eur. Phys. J. Plus}  {\bf 136}, 976 (2021).


	
\end{thebibliography}
\end{document}